\documentclass[aps,prl,twocolumn,showpacs,superscriptaddress,groupedaddress]{revtex4}  
\usepackage{graphicx}  
\usepackage{dcolumn}   
\usepackage{bm}        
\usepackage{amssymb}   

\begin{document}

\hspace{5.2in} \mbox{Fermilab-Pub-04/xxx-E}

\title{Monofractal nature of air temperature signals reveals their climate variability}

\author{A.~Deli\`ege} \affiliation{University of Li\`ege, Li\`ege, Belgium}
\author{S.~Nicolay} \affiliation{University of Li\`ege, Li\`ege, Belgium}

\vskip 0.25cm

\date{\today}

\begin{abstract}
We use the discrete ``wavelet transform microscope'' to show that the surface air temperature signals of weather stations selected in Europe are monofractal. This study reveals that the information obtained in this way are richer than previous works studying long range correlations in meteorological stations. The approach presented here allows to bind the H\"older exponents with the climate variability. We also establish that such a link does not exist with methods previously carried out. 
\end{abstract}

\pacs{05.45.Tp, 02.30.Sa, 05.45.Df}
\maketitle


Previous works have shown the presence of long range correlations (LRC) in the trend/noise of surface air temperature data (see e.g. \cite{koscielny,koscielny96,bunde}). If such a signal is interpreted as a random walk, one can conclude to its monofractal nature. Since these studies have been performed with ``bias-dependent'' methods, they can not be directly applied to the ``raw signal'': so-called seasonal variations have first to be removed (\cite{arneodo95,koscielny}). The purpose of this letter is to show that raw temperature signals contain more information than the associated trends. We first present the wavelet leaders method (WLM) as a tool for providing a multifractal formalism, which has proven to be well suited to study fractal objects (\cite{abry,jaffard04,jaffard09,  lashermes,wendt}). We then use this wavelet-based approach to show that surface air temperature signals are monofractal. Finally, we show that the fluctuation of the monofractal exponent observed from one station to another is bonded to the climate variability. Such a relation is not observed with ``bias-dependent'' methods.

%
Let us first recall the WLM. The discrete wavelet transform (WT) allows to decompose a signal in terms of wavelets that are constructed from a single function $\psi$ (\cite{daubechies,mallat}). The WT of a function $f$ is defined as
\[
 W_\psi[f](j,k)= 2^{-j} \int f(x) \psi (2^{-j} x +k) dx,
\]
where $k$ is the space parameter and $j$ the scale parameter (both take integer values). WT is well adapted to study the irregularities of $f$, even if they are masked by a smooth behavior. If $f$ has, at a given point $x_0$, a local scaling/H\"older exponent $h(x_0)$, in the sense that $|f(x)- P_{x_0}(x)|\sim |x-x_0|^{h(x_0)}$ around $x_0$, where $P_{x_0}$ is a polynomial of degree at most $h(x_0)$, then with the right choice of $\psi$, one has $W_\psi[f](j,k)\sim 2^{-jh(x_0)}$ for the indices $k$ such that $2^{-j}x-k$ is close to $x_0$ (\cite{jaffard04,jaffard09}). The WLM is a transposition of the wavelet transform modulus maxima (WTMM) to the discrete setting (\cite{arneodo95-2,jaffard04,jaffard06,jaffard09}). Mimicking the box-counting technique, one investigates the scaling behavior of the following partition function
\[
 S(q,j) = 2^j \sum_{k} (\sup_{j'\ge j}|W_\psi[f](j',k)|)^q \sim 2^{j \omega(q)},
\]
where $q$ is a real number. In this framework, $\omega$ is the Legendre transform of the singularity spectrum, defined as the Hausdorff dimension of the set of points $x$ sharing the same H\"older exponent $h(x)$. Monofractal functions, i.e.\ functions with a constant H\"older exponent $h(x_0)=H$ are characterized by a linear spectrum: $H=\partial \omega/\partial q$. On the contrary, a nonlinear $\omega$ curve is the signature of functions displaying a multifractal behavior; in this case, $h$ is not constant anymore and thus may fluctuate from one point to another.

\begin{figure}
\begin{center}
\includegraphics[scale=0.12]{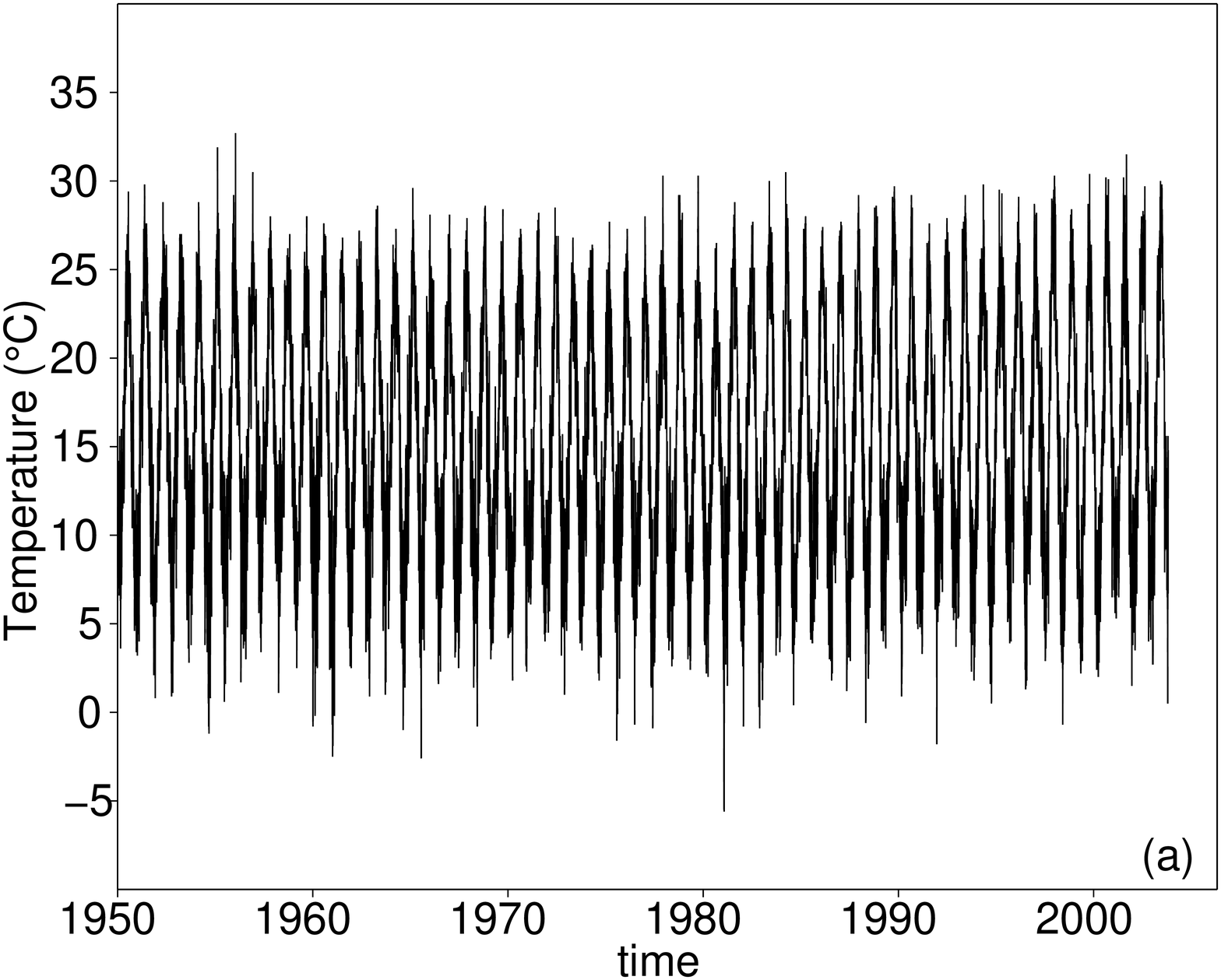} \includegraphics[scale=0.12]{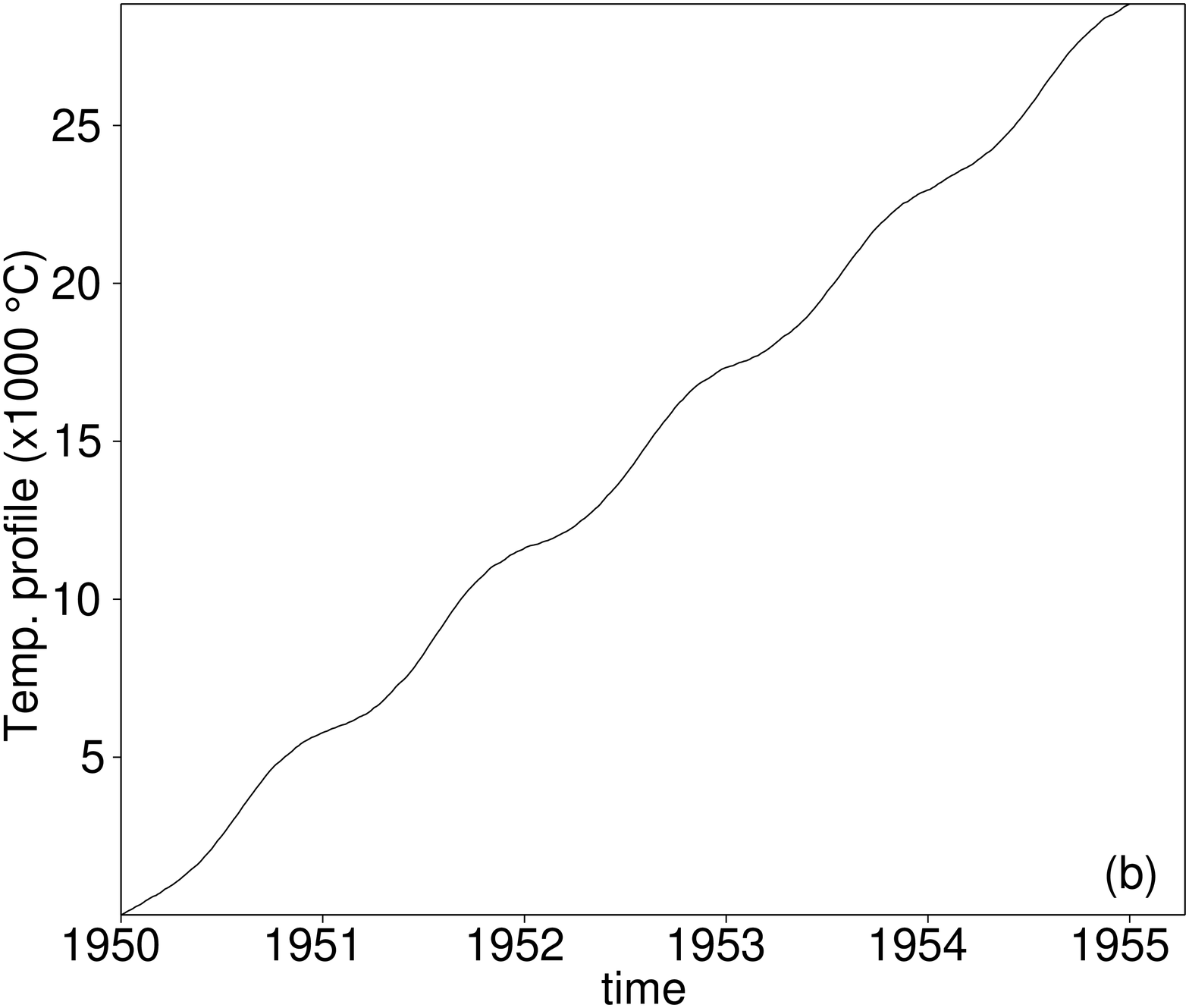} \\ \includegraphics[scale=0.12]{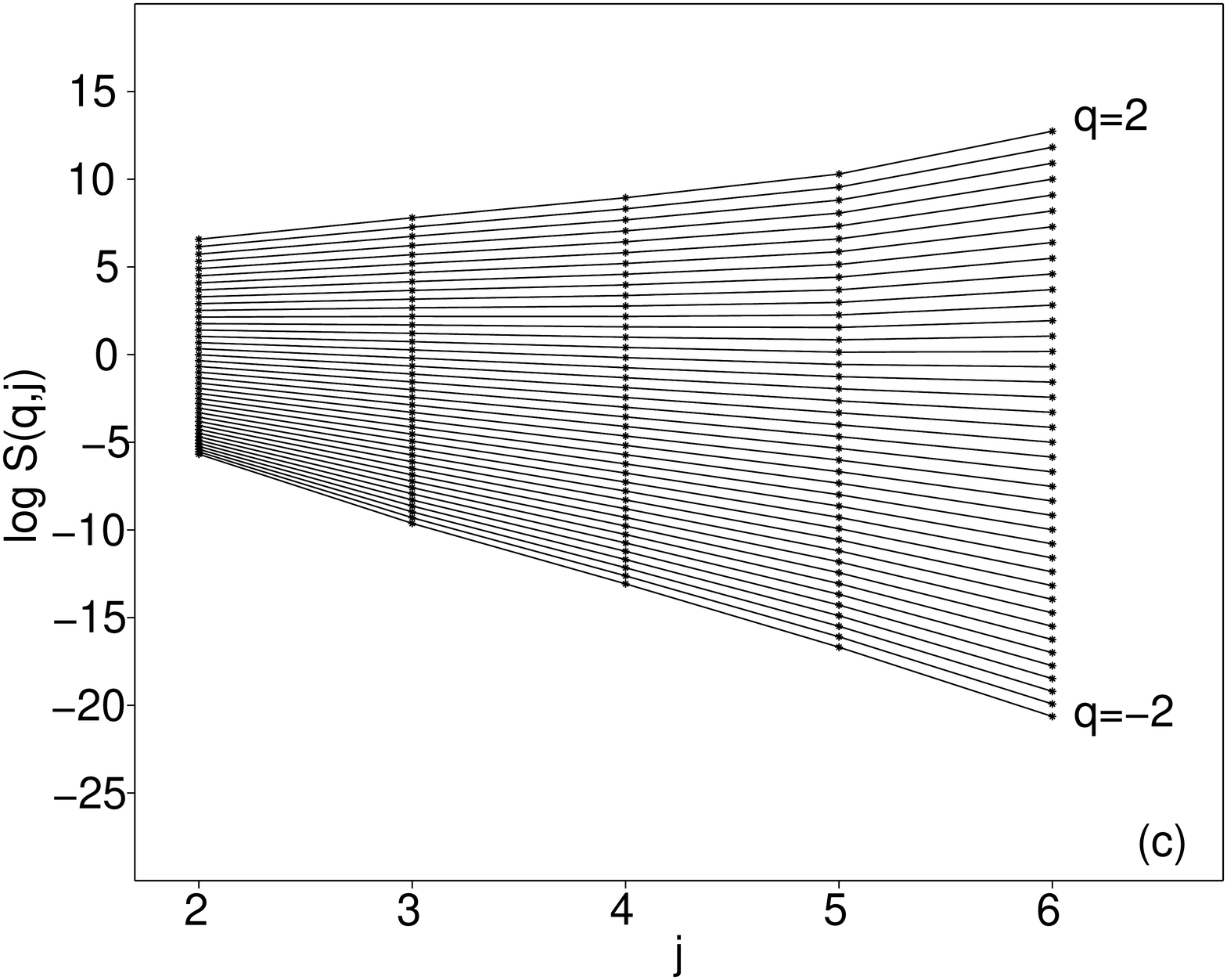} \includegraphics[scale=0.12]{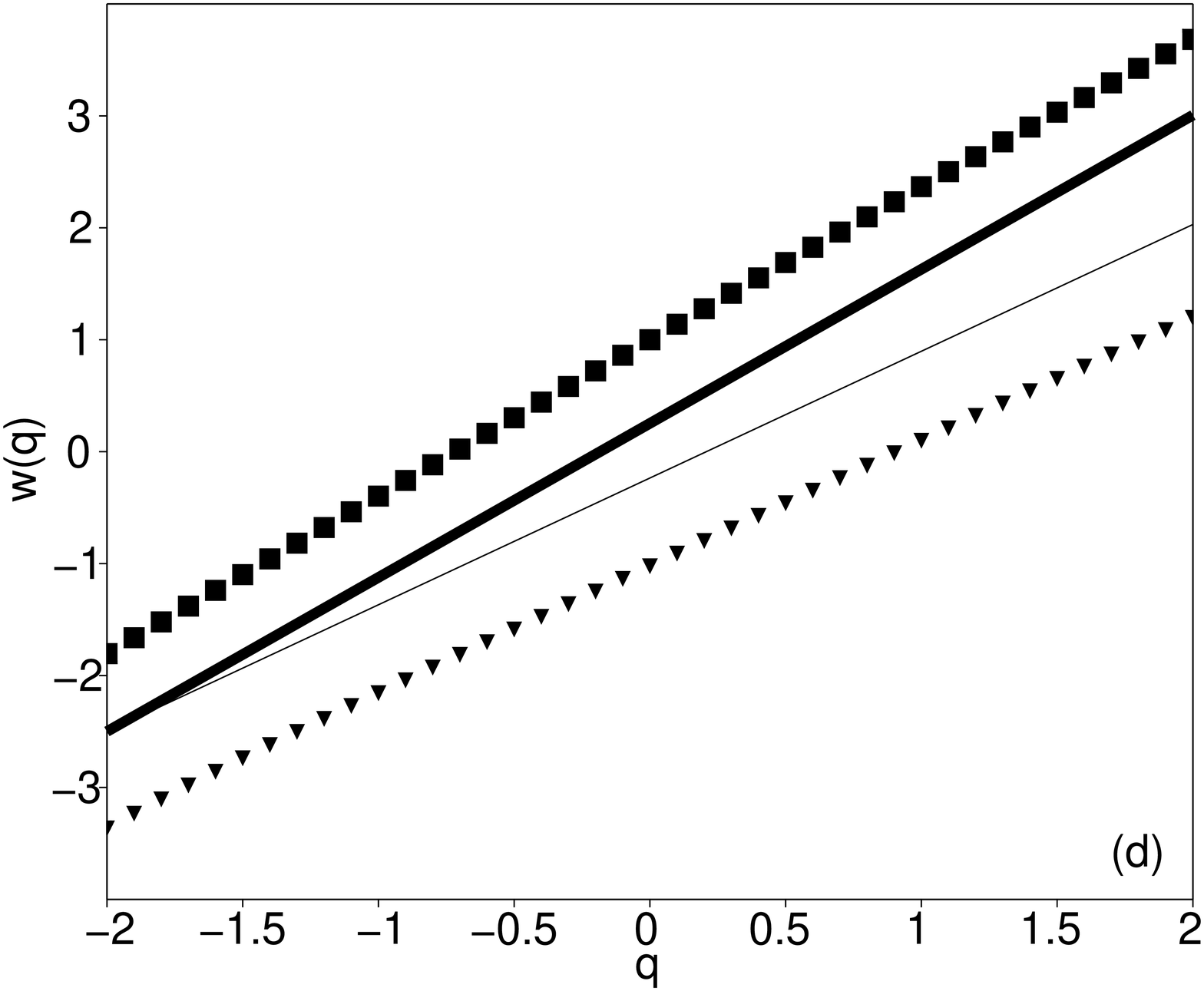}
\end{center}
\caption{(a) Raw signal of a weather station located in Rome, Italy, from 1950 to 2003. (b) Temperature profile of Rome from 1950 to 1955. (c) $\log S(q,j)$ vs $j$ for $q$ ranging from $-2$ to $2$ (from bottom to top) by step of $0.1$. For a fixed $q,$ the slope of the linear regression over $\log S(q,j)$ gives the value of $\omega(q)$ (see (d)). (d) Comparison of the functions $\omega(q)$ for Rome (squares) and for Armagh (Ireland, triangles). The thick straight line represents the linear regression line of $\omega$ corresponding to Rome, the other one corresponds to Armagh. Both functions $\omega$ are clearly linear, which implies that the signals are monofractal with H\"older exponents given by the slopes of the regression lines: $1.38$ for Rome and $1.13$ for Armagh.}
\label{fig:signaux}
\end{figure}

We applied the WLM on surface air temperature data collected from \cite{ecad}. In order to get homogeneous signals, we limited our study to daily mean temperature series with at least 50 years of data between 1950 and 2003 spread across Europe between $36^\circ$ (Southern Spain, Italy, Greece) and $55^\circ$ of latitude (Northern Ireland, Germany) and $-10^\circ$ (Western Ireland, Portugal) and $40^\circ$ of longitude (Eastern Ukraine). By doing so, we were able to select 115 weather stations uniformly dispersed across the selected area. For the purpose of reducing the noise, the data $f(t)$ were replaced with the temperature profiles $\sum_{u=1}^t f(u)$ (see Fig. \ref{fig:signaux}(a) and (b)). All the air temperature data display a monofractal nature: every function $\omega$ is clearly linear with a mean coefficient of determination equal to $R^2=0.9975 \pm 0.0028$ (see Fig. \ref{fig:signaux}(c) and (d)). However, the value of the H\"older exponent varies from one station to another between $1.093$ and $1.43$ (see Fig. \ref{fig:signaux}(d)). Let us also remark that other ``bias-independent'' methods give similar results for each station; we performed the WTMM (\cite{arneodo95-2}) as well as the $S^\nu$-based multifractal formalism (\cite{kleyntssens}) on the data to confirm our results.

\begin{figure}
\centering
\includegraphics[scale=0.12]{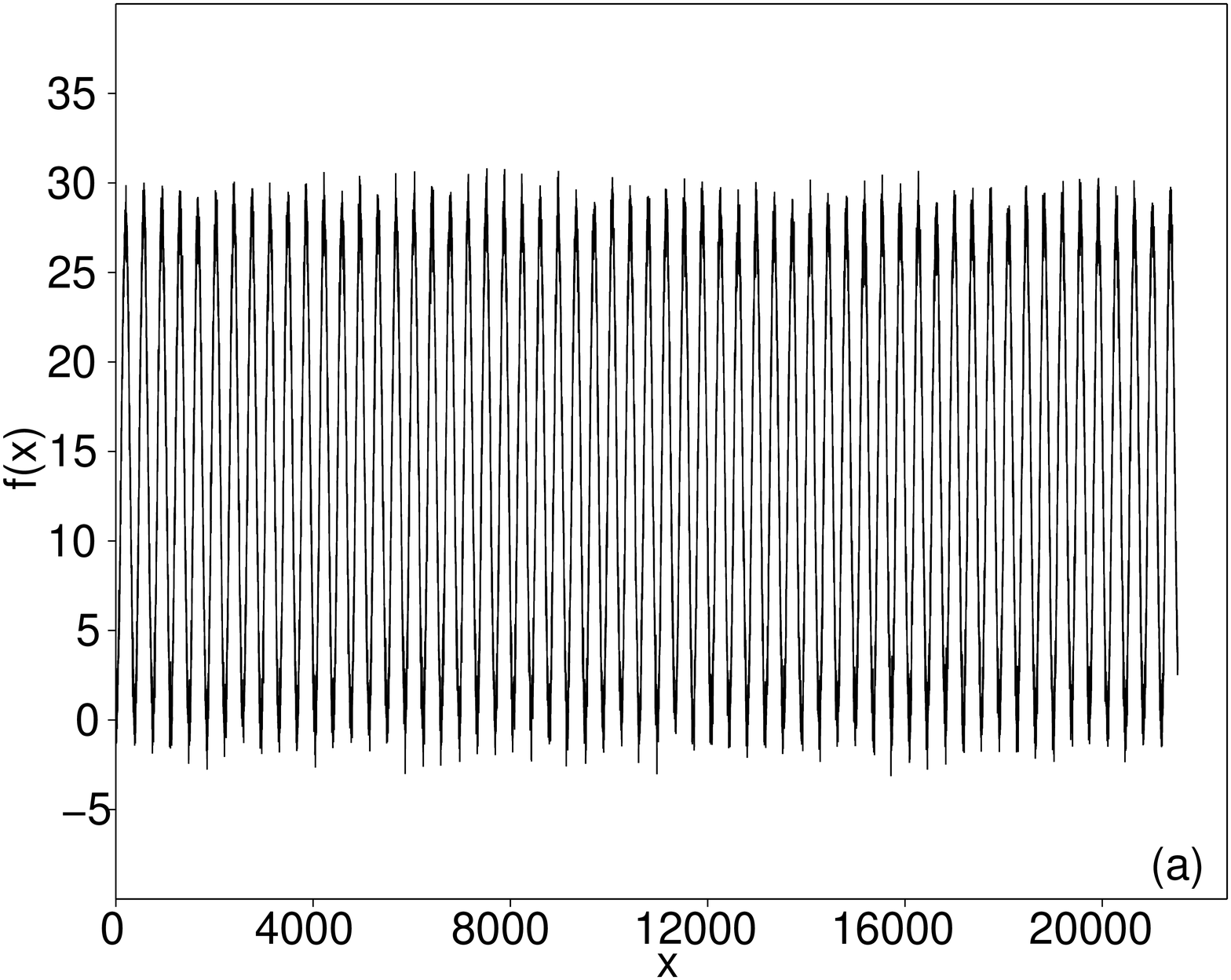} \includegraphics[scale=0.12]{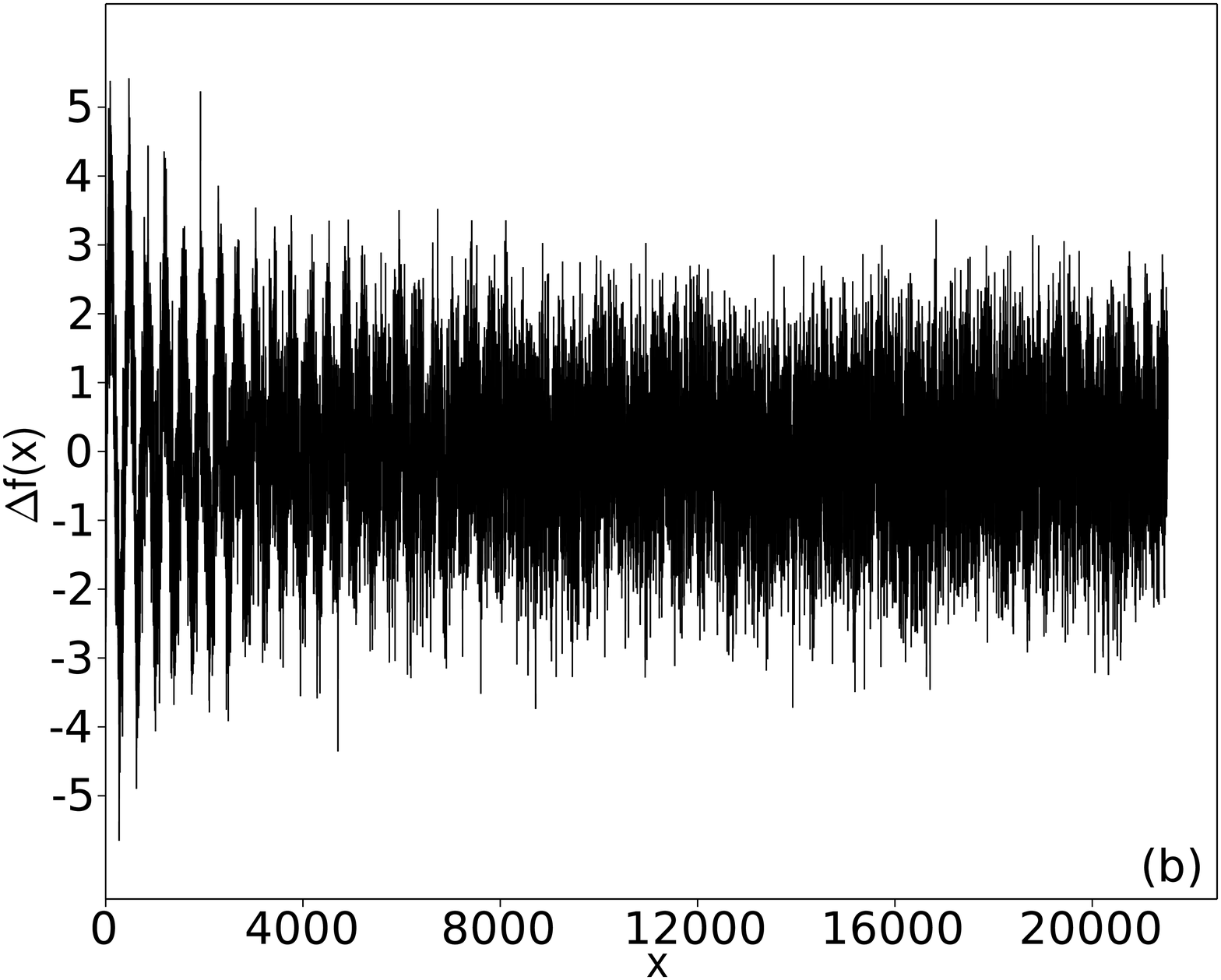} \\ \includegraphics[scale=0.12]{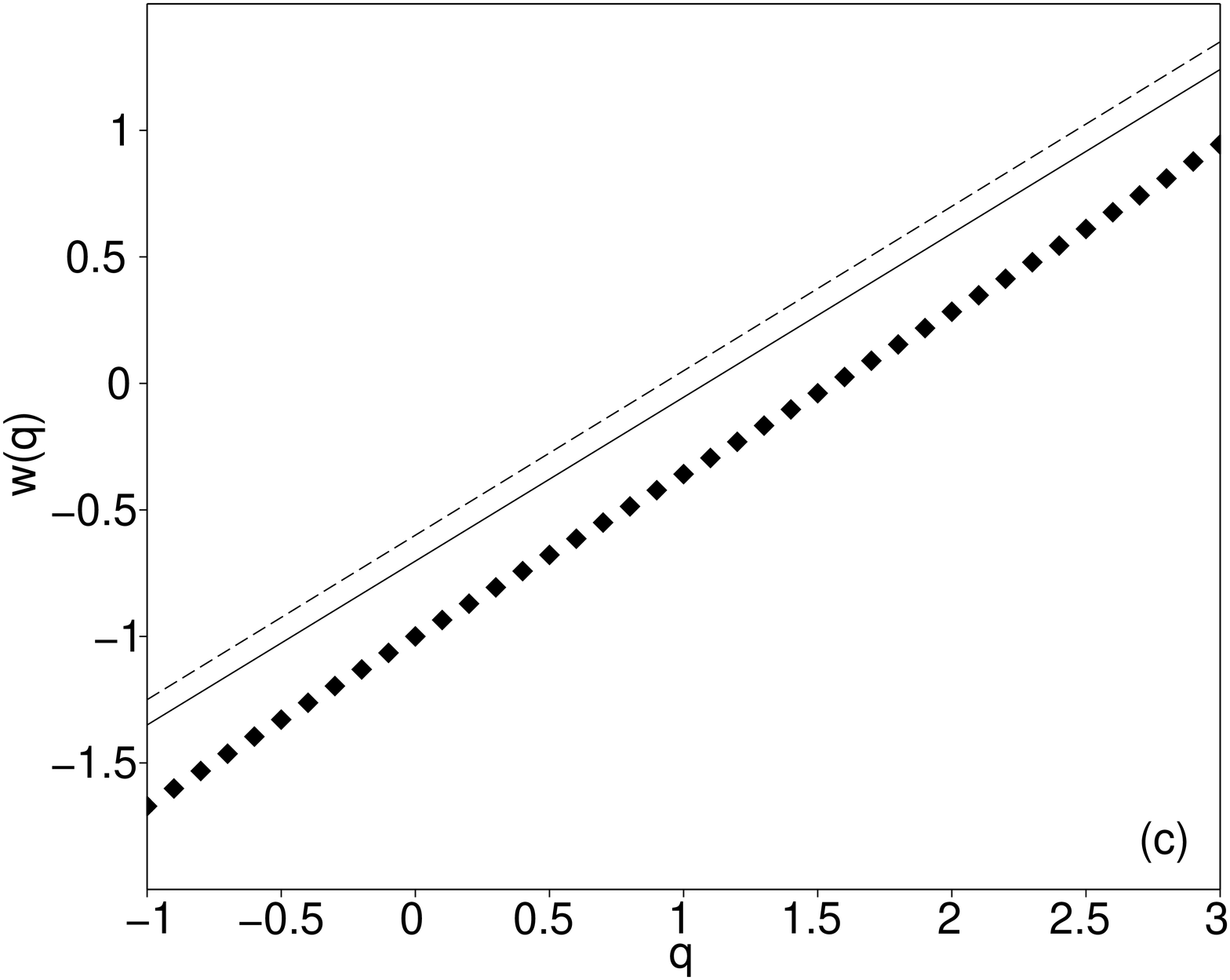} \includegraphics[scale=0.12]{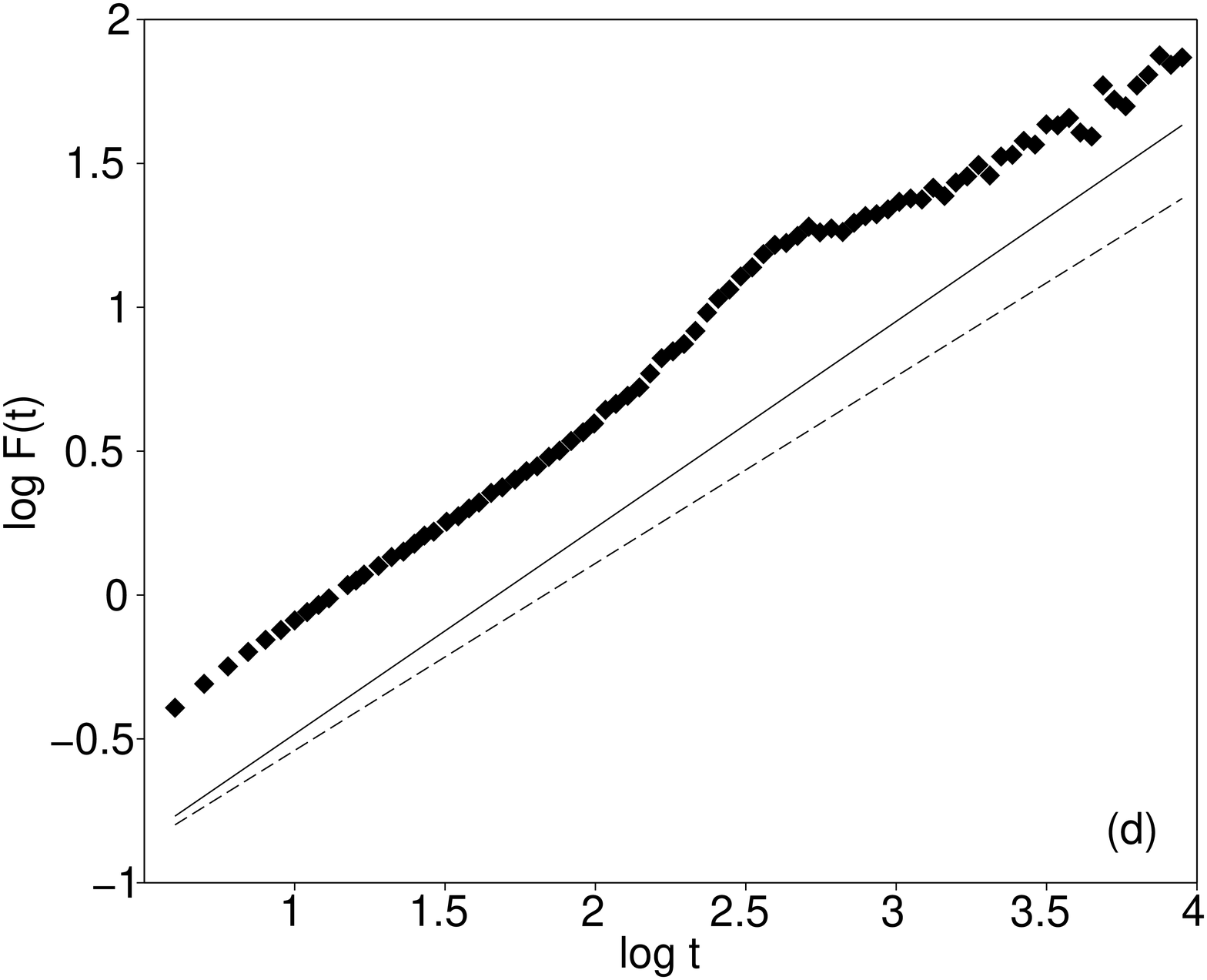}
\caption{(a) A synthetic signal mimicking temperature data made of a fractional Gaussian noise with LRC with index 0.65 and a non-stationary seasonal variation (see text). The signal is therefore monofractal with H\"older exponent 0.65. (b) The corresponding trend of the signal, i.e. $\Delta f(x) = f(x) - \left\langle f\right\rangle (d)$ where $\left\langle f\right\rangle (d)$ is the mean periodic variation (see text). (c) Function $\omega(q)$ obtained from WLM (diamonds). The straight line corresponds to the linear regression of $\omega$, which has slope 0.648, and the dashed line is the line with the expected slope 0.65. Clearly WLM gives accurate results. (d) Log-log plot of the standard deviation $F$ obtained from DFA (diamonds). One can clearly see that $\log F$ is not quite linear. The straight line is the ``best'' linear regression line of $ \log F$, which has slope 0.717, whereas the expected value is 0.65 (dashed line).}
\label{fig:contrex}
\end{figure}

Influential studies about the monofractal nature of air temperature data have been previously carried out (see e.g. \cite{bunde}, \cite{koscielny96}, \cite{koscielny}). However, the approach adopted here fundamentally differs for one reason: we apply the WLM on raw data, which is not possible with ``bias-dependent methods'' such as the detrended fluctuation analysis (DFA) used in \cite{koscielny}. For the sake of comparison, let us briefly describe the DFA first introduced by in \cite{peng94},\cite{peng95}. Following \cite{koscielny}, if $f$ is a surface air temperature time series, the seasonal variation $\langle f\rangle$ is defined as follows: if $d$ is a calendar date (say June first), $\langle f\rangle(d)$ is the average over the years in $f$ of the values $f(t)$ such that $t$ corresponds to the calendar date $d$ (June first 1950, June first 1951,...). The corresponding trend is then defined as $\Delta f(t)= f(t)- \langle f\rangle (d)$. 
To reduce the noise, the temperature profiles $\sum_{u=1}^t \Delta f(u)$ are also used instead of the trend. From random walk theory (see e.g. \cite{barabasi}), the standard deviation $F$ of the profile in a time window of length $t$ should behave like $F(t)\sim t^{\gamma}$, where $\gamma>1/2$ suggests the existence of LRC and is the H\"older exponent of the data. For the DFA, the best linear fit is determined on every non-overlapping segment $\eta$ of length $t$ and the standard deviation $F_{\eta}(t)$ of the profile from that straight line is then computed. Finally, $F(t)$ is defined as $\sqrt{E[F^2_{\eta}(t)]}$, where $E$ stands for the mean over all segments. By doing so, one gets rid of the influence of the possible linear trends on scales larger than $t$.
%
As a simple example, if one considers a signal $f$ made of a sine (representing the seasonal variation) and a fractional Gaussian noise (FGN) \cite{mandelbrot}, both methods will match, i.e.\ will detect the monofractality of the FGN. The same result is obtained if one applies the WLM on the signal where the seasonal variation has been removed (thus proceeding in the same way as for the DFA). However, this concordance is not recovered if one applies these methods on real surface air temperature time series. In this case, DFA and WLM with seasonal variation removal give similar results, while the WLM applied on the raw data leads to different outcomes. This is because information remains in the seasonal variation. This can be illustrated with a synthetic signal roughly mimicking temperature data. Let $n$ be a FGN associated to LRC with index $\gamma=0.65$,
\[
 s(x) =15 \sin \left(\frac{2\pi}{365} x -\frac{\pi}{2} -\frac{1}{20} \log(x+1) \right) +14
\]
be a non-stationary seasonal variation and define $f(x)= n(x) +s(x)$ (see Fig. \ref{fig:contrex}(a) and (b)). DFA applied to $f$ does not lead to a straight line (see Fig. \ref{fig:contrex}(d)) and the best result for the estimation of $\gamma$ we can hope for (knowing the expected value of $\gamma$) is an error of order $10^{-1}$. On the other hand, the WLM applied on $f$ works properly (see Fig. \ref{fig:contrex}(c)), giving a H\"older exponent equal to $\gamma$ with an error of order $10^{-3}$.

%
%
A natural question arising is whether or not the observed H\"older exponent reflects the climate variability. More precisely, does the surface pressure anomalies induce differences in the H\"older exponents, or are these differences numerical artifacts? To test this hypothesis, we compared the map of the surface pressure anomalies from \cite{ncep} with the same map where the anomalies have been replaced with the measured H\"older exponent. On these maps, each pixel, corresponding to an anomaly or a H\"older exponent (both related to a weather station), is renormalized in order to obtain values between $0$ and $1$. One can compute the Frobenius distance between two such maps (considered as matrices) as follows:
\[
 d=\sqrt{\sum_{i,j} (x_{i,j}- x'_{i,j})^2},
\]
where $x_{i,j}$ is a pixel of the first map, $x'_{i,j}$ is the corresponding pixel of the second map and where the sum is taken over all pixels. In this case, the distance between these two maps is $d_1 = 2.68$.
\begin{figure}
\centering
\includegraphics[width=0.7\columnwidth, height=4cm]{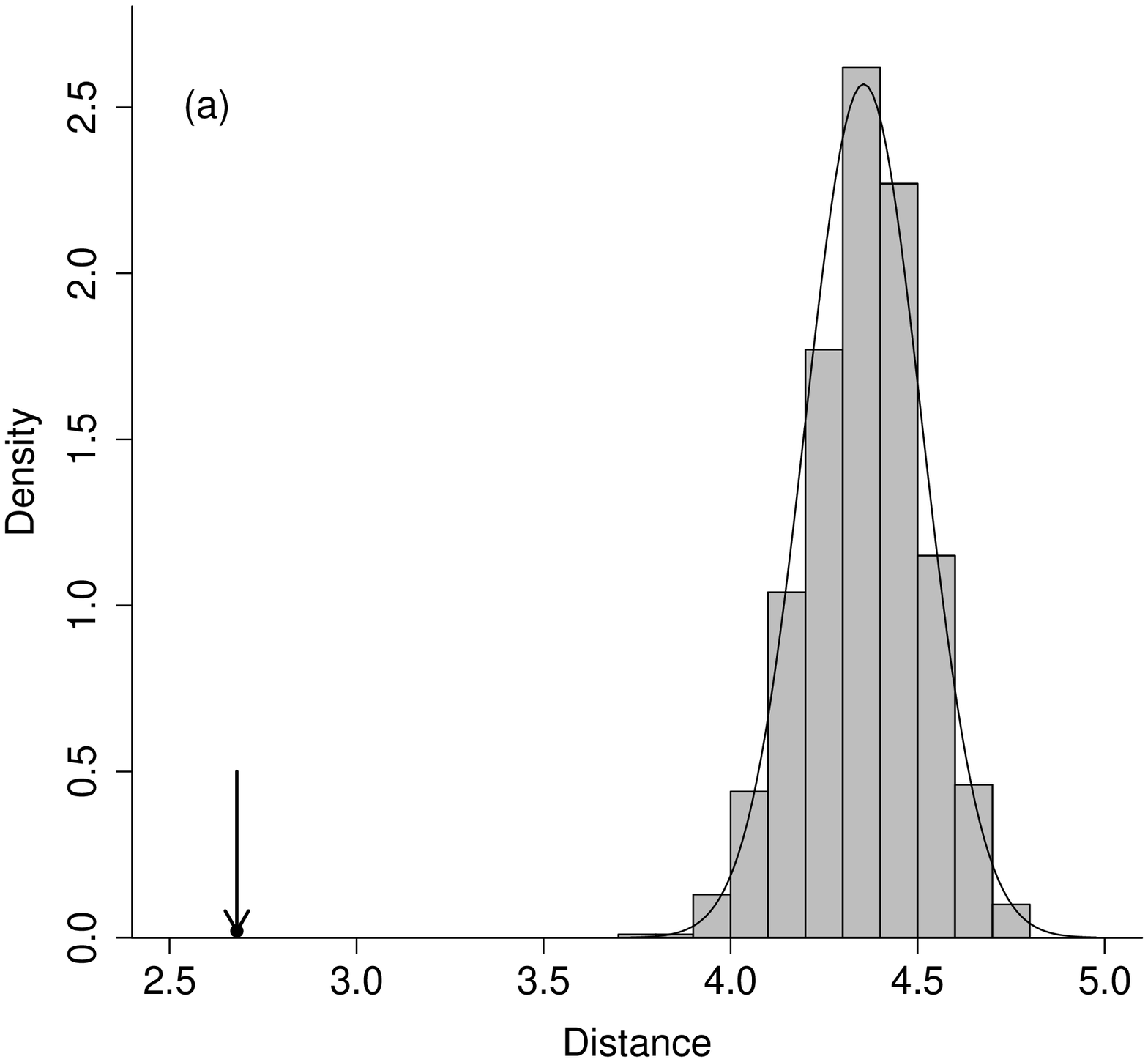} \\ \includegraphics[width=0.7\columnwidth, height=4cm]{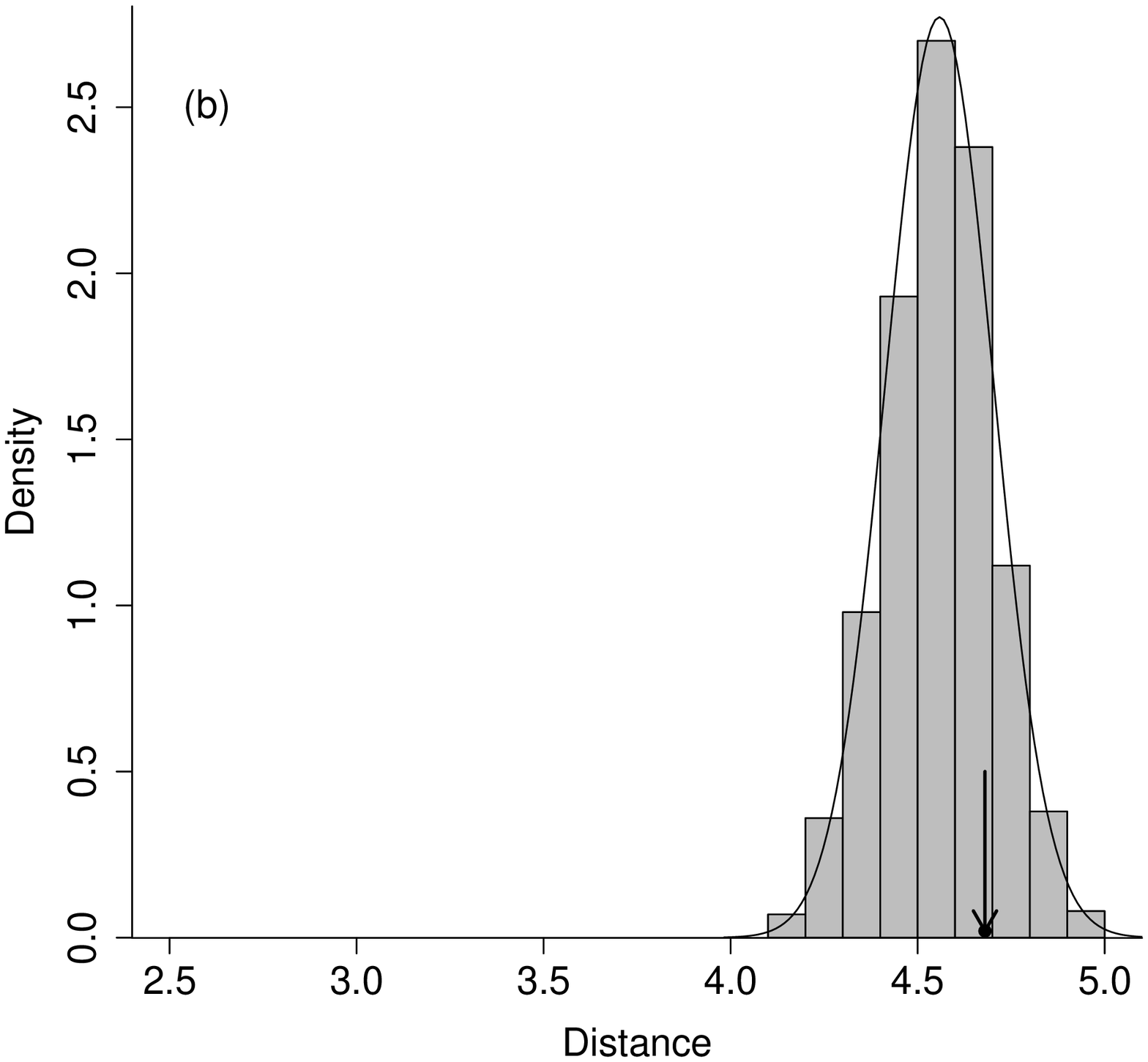}
\caption{(a) (resp.\ (b)): histogram of the distances between the randomly shuffled H\"older maps (resp.\ randomly shuffled DFA maps) and the pressure anomalies map. The curves represent the theoretical Gaussian distributions based on the mean and the standard deviation of the measured distances. The arrow indicates the distances $d_1$ (resp.\ $d_2$) between the non-shuffled H\"older map (resp.\ DFA map) and pressure anomalies map, i.e. $d_1 = 2.68$ (resp.\ $d_2 = 4.68$). Obviously, original H\"older map and pressure anomalies map are close, in the sense that it is extremely unlikely to measure a distance smaller than $2.68$ if the H\"older map is randomly shuffled. On the opposite, exponents obtained with DFA do not seem to be related to pressure anomalies since the distance between them is not affected by the shuffling.}
\label{fig:hist}
\end{figure}
To check if $d_1$ is ``large'', the ``H\"older map'' was randomly shuffled $10,000$ times. For each realization, the distance with the original anomalies map was computed in order to get a distribution of these random distances. In this way, one can look where $d_1$ lies in the distribution of the distances, and one can associate a $p$-value to this particular distance $d_1$. Based on the $10,000$ observations, the probability $1-p$ to have a randomly shuffled map with a distance smaller than $d_1$ is lower than $10^{-4}$, which shows that the H\"older map and the pressure anomalies map are highly significantly close (see Fig. \ref{fig:hist}(a)). In order to show that other methods do not give so good results, we performed the same simulation but with a map where the H\"older exponents obtained with the WLM have been replaced with the values obtained with the DFA. In this case, the distance $d_2$ between this ``DFA map'' and the anomalies map is $4.68,$ and the probability that the distance between a randomly shuffled DFA map and the anomalies map is smaller than $d_2$ is $1-p= 0.8$. This shows that the DFA map has to be considered as random (see Fig. \ref{fig:hist}(b)). One can thus conclude that the H\"older exponents obtained via the DFA have no obvious relation with the climate variability.

As a conclusion, one can say that the trend/noise studied in \cite{koscielny} is monofractal but uniform, while the whole signal is also monofractal but not uniform. Moreover, H\"older exponent obtained here with the WLM reflects the climate variability of the station associated to the data, which is not the case with ``bias-dependent'' methods.

\bibliographystyle{plain}
\bibliography{Biblio}

\begin{thebibliography}{10}

\bibitem{abry}
P.~Abry, H.~Wendt, S.~Jaffard, H.~Helgason, P.~Goncalves, E.~Pereira,
  C.~CHarib, P.~Gaucherand, and M.~Doret.
\newblock Methodology for multifractal analysis of heart rate variability: From
  lf/hf ratio to wavelet leaders.
\newblock In {\em Nonlinear Dynamic Analysis of Biomedical Signals EMBC
  conference (IEEE Engineering in Medicine and Biology Conferences)}, 2010.

\bibitem{arneodo95}
A.~Arneodo, E.~Bacry, P.V. Graves, and J.F. Muzy.
\newblock Characterizing long-range correlations in dna sequences from wavelet
  analysis.
\newblock {\em Phys. Rev. Lett.}, 74 (16):3293--3297, 1995.

\bibitem{arneodo95-2}
A.~Arneodo, E.~Bacry, and J.F. Muzy.
\newblock The thermodynamics of fractals revisited with wavelets.
\newblock {\em Physica A}, 213:232--275, 1995.

\bibitem{ecad}
European~Climate Assessment and Dataset.
\newblock \url{http://eca.knmi.nl}.

\bibitem{barabasi}
A.-L. Barabasi and H.E. Stanley.
\newblock {\em Fractal Concepts in Surface Growth}.
\newblock Cambridge University Press, 1995.

\bibitem{bunde}
A.~Bunde and S.~Havlin.
\newblock Power-law persistence in the atmosphere and in the oceans.
\newblock {\em Physica A}, 314:15--24, 2002.

\bibitem{daubechies}
I.~Daubechies.
\newblock {\em Ten lectures on Wavelets}.
\newblock SIAM, 1992.

\bibitem{jaffard04}
S.~Jaffard.
\newblock Wavelet techniques in multifractal analysis.
\newblock {\em Proceedings of symposia in pure mathematics}, 72:91--152, 2004.

\bibitem{jaffard06}
S.~Jaffard, B.~Lashermes, and P.~Abry.
\newblock {\em Wavelet Analysis and Applications}, chapter Wavelet leaders in
  multifractal analysis, pages 201--246.
\newblock Birkauser, 2006.

\bibitem{jaffard09}
S.~Jaffard and S.~Nicolay.
\newblock Pointwise smoothness of space-filling functions.
\newblock {\em Appl. Comput. Harmon. Anal.}, 26:181--199, 2009.

\bibitem{kleyntssens}
T.~Kleyntssens, C.~Esser, and S.~Nicolay.
\newblock A multifractal formalism based on the {$S^\nu$} spaces: From theory
  to practice.
\newblock Submitted for publication.

\bibitem{koscielny96}
E.~Koscielny-Bunde, A.~Bunde, S.~Havlin, and Y.~Goldreich.
\newblock Analysis of daily temperature fluctuations.
\newblock {\em Physica A}, 231:393--396, 1996.

\bibitem{koscielny}
E.~Koscielny-Bunde, A.~Bunde, S.~Havlin, H.E. Roman, Y.~Goldreich, and H.-J.
  Schellnhuber.
\newblock Indication of a universal persistence law governing atmospheric
  variability.
\newblock {\em Phys. Rev. Lett.}, 81:729--732, 1998.

\bibitem{lashermes}
B.~Lashermes, S.G. Roux, P.~Abry, and S.~Jaffard.
\newblock Comprehensive multifractal analysis of turbulent velocity using
  wavelet leaders.
\newblock {\em Eur. Phys. J. B.}, 61 (2):201--215, 2008.

\bibitem{mallat}
S.~Mallat.
\newblock {\em A Wavelet Tour of Signal Processing}.
\newblock Academic Press, 1999.

\bibitem{mandelbrot}
B.B. Mandelbrot and J.W.~Van Ness.
\newblock Fractional brownian motions, fractional noises and applications.
\newblock {\em SIAM}, 10 (4):422--437, 1968.

\bibitem{peng94}
C.-K. Peng, S.V. Buldyrev, S.~Havlin, M.~Simons, H.E. Stanley, and A.L.
  Goldberger.
\newblock Mosaic organization of dna nucleotides.
\newblock {\em Phys Rev E}, 49:1685--1689, 1994.

\bibitem{peng95}
C.-K. Peng, S.~Havlin, H.E. Stanley, and A.L. Goldberger.
\newblock Quantification of scaling exponents and crossover phenomena in
  nonstationary heartbeat time series.
\newblock {\em Chaos}, 5:82--87, 1995.

\bibitem{ncep}
NCEP/NCAR Reanalysis.
\newblock \url{http://www.esrl.noaa.gov/}.

\bibitem{wendt}
H.~Wendt, P.~Abry, S.~Jaffard, H.~Ji, and Z.~Shen.
\newblock Wavelet leader multifractal analysis for texture classification.
\newblock In {\em Proc IEEE conf. ICIP}, 2009.

\end{thebibliography}

\end{document}